\pageno=0
\nopagenumbers
\footline= { \ifnum\pageno>0 \hss\tenrm\folio\hss \fi}
\magnification=\magstep1

\hsize=6truein
\vsize=8.5truein
\null 
\tenrm 
\def \i{{\rm i}}
\def \e{{\rm e}}
\def \um{{\textstyle {1\over2}}}
\def \uq{{\textstyle {1\over4}}}
\def \uo{{\textstyle {1\over8}}}
\def \dte{{\textstyle {2\over3}}}
\def \udp{{\textstyle {1\over{2\pi}}}}
\def \B {|\circ >}
\def \BB {<\circ }
\def \S {|\bullet>}
\def \SS {<\bullet}

\centerline{\bf A HAMILTONIAN MODEL }
\centerline{\bf FOR HARD INELASTIC HADRONIC COLLISIONS }

\vskip 2pc

\centerline{G. CALUCCI and D. TRELEANI}

\centerline{{\it Dipartimento di Fisica teorica dell'Universit\`a di Trieste,
 I 34014}}
\centerline{{\it INFN, Sezione di Trieste, Italy}}
\vskip 3pc

{\midinsert

\centerline{\bf {Abstract} }
\narrower\narrower
A Hamiltonian eikonal model for multiple production in high energy 
hadron-hadron collisions is presented and worked out with the aim of providing 
a
simple
frame for various different observables.
\par
An important role is played by unitarity which is built in by construction in
the Hamiltonian formulation. The
eikonal approximation allows both a very effective simplification of the
dynamics, and facilitates the discussion on the relevance of possible spatial
inhomogeneities of the hadrons.
The model is intended to describe only the hard interaction of the 
constituents,
the structure of the incoming hadrons and the final hadronization processes are
outside the scope of the present investigation.
\endinsert}

\vfill
\eject 

{\bf 1. Introduction}
\vskip 1pc
A Hamiltonian model for the description of the multiple production process in
hadron-hadron collision is presented with the main aim of bringing together
different observables within a unique frame. A particular attention is given to
those features of the inelastic processes that can give informations on the
proton structure. among these the relation between the hard-inelastic cross 
section and the inclusive production rates. 
\par
The physical ingredients of the model are the following:
\item{ $\bullet$}
Although the underlying theory should be ideally QCD there is a
sharp distinction between soft dynamics, which provides the binding of the
partons in the hadrons and also the final hadronization of the shaken-off 
partons, and hard dynamics that causes the parton scattering.
\item{ $\bullet$}
Hard collisions gives a finite transverse momentum to the partons, which
remains however small with respect to the typical longitudinal momentum.
Hard rescattering is included, but not hard branching of the partons.
\item{ $\bullet$}
Discrete quantum numbers like spin and colour are not taken into account.
\par
 The observable quantities that can be computed
are the hard inelastic cross section, the inclusive cross section for the 
production
of
back-to-back pairs of partons, the cross section for double pair production,
the multiplicity distribution, the backward forward correlation between the 
produced
partons. The effect on the observables of the hard dynamics, described by 
the Hamiltonian, will be distinguishable from the effect of 
the hadronic
structure, which is parametrized in an independent way.
\par
In the second chapter, after a short reminder of the eikonal formalism
[1], whose use is suggested by the kinematical conditions, the general features
 of the model are made precise by
defining the interaction and by choosing a definite partonic description of
the hadron.
In the third chapter the expression for the inelastic cross section is derived.
In the fourth chapter a few observables related to the production process are
calculated.
In the fifth chapter the possibility of a non uniform distribution of the 
partonic matter in the hadron is considered and the effects of this hypothesis
on the inelastic cross section and on the production rates are worked out.
Since the whole treatment deals with the transverse variables an exploration 
on
the possibility of taking, somehow, into account the longitudinal degrees of
freedom and of some simple related consequences is presented in the sixth 
chapter where
also an initial discussion of more general forms of parton distribution inside
the hadron is sketched.
\par 
One can find lot of previous treatment sharing important analogies with the
 treatment presented here, both in the eikonal formulations for the multiple
production and in some purely probabilistic descriptions of the collision 
processes, which however, providing a form of unitarization of the transition
probabilities, are in their final answer, analogous to the present
formulation[2]. The main advantage in the actual approach with the Hamiltonian
formalism is that unitarity is explicitly implemented in all the different 
steps
of the calculation and its role, with respect to the different observables
considered, can be always traced back.
\par
  A preliminary version of this work was presented at the {\it XXVIII 
  International Symposium on Multiparticle Dynamics - Delphi (1998)}[3].
 
\vskip 2pc
{\bf 2. General features}
\vskip 1pc
{\it 2.1 A short reminder of the eikonal approximation}
\vskip 1pc

The description and justification of the eikonal formalism in high energy
scattering has been presented in a lot of papers, so there is no point in
re-deriving it. Only some features that are relevant for the next exposition 
are
here briefly recalled [1,4].
\par
The relative motion of two very fast colliding particles in their c.m. frame is
described by the free Hamiltonian;
$$H_o={\bf v\cdot p}+M/\gamma$$
One then adds an interaction term $V$ which is for the moment left
unspecified, but for the fact that it depends on the relative coordinate 
${\bf r}=({\bf B},z)$. At very high energy the speed remains practically 
constant even
for sizable changes in the momentum, so the solution of this Hamiltonian 
problem is given by the wave-function:
$$\Psi({\bf r})={1\over{(2\pi)^{3/2}}}
\exp\Bigl[- {{\i}\over\beta}\int^z_{-\infty}dz'V({\bf B},z')+\i 
pz\Bigr]\eqno(2.1)$$
The exponential factor yields, in the limit $z\to\infty$, the
S-matrix at fixed $B$.
>From this formula it appears clearly that the most important feature of the
potential is its dependence on $B$ whereas its longitudinal variables are 
always
integrated; moreover the integration in $dz'/\beta$ is equivalent to an
integration over time.
\par
The considerations here presented are of pure kinematical origin, they keep 
their
validity also when a more complicated structure is foreseen both for the 
interaction term and for the incoming states as it will be done in the next 
section. The main improvement will be the description of the incoming states as
systems with internal degrees of freedom [5,6] and thus the introduction of a 
set of transverse coordinates, one for each component.
\vskip 1pc
{\it 2.2 Description of the model}
\vskip 1pc

The model describes the hadrons as sets of bound partons which, due to the
interaction, may become finally free and eventually may be detected as jets in 
the
final state; the hadronization
process is not described.
\par
The only detailed kinematics is the transverse one, the longitudinal is in some
way integrated over, so also the longitudinal relative motion of the hadrons, 
which
appears in eq.(2.1) as $\e^{\i pz}$, is not explicitly written out. 
 Since we deal with very-high energy collisions there
is a sharp distinction between backward and forward degrees of freedom.
We call $a_b,a_f$ the operators of the bound backward and forward partons and 
$c_b,c_f$ the operators of the free partons. They have the standard 
commutation relations, every backward operator commutes with every forward
operator and every $a$ commutes with every $c$;they are local in ${\bf b}$, the
transverse impact parameter of the parton, this is possible since the size of
the region relevant for the hard scattering is much smaller than the hadron 
size;
so we can write a free Hamiltonian:
$${\cal H}_o=\sum_{v=f,b} \omega \int d^2
b[a^{\dagger}_v({\bf b})a_v({\bf b})+c^{\dagger}_v({\bf b})c_v({\bf b})]
\eqno(2.2)$$
The interaction that we want to describe is the hard collision of two bound
partons that give rise to two free partons is such a way however that they keep
their property of being either backward or forward. Thus the interaction 
Hamiltonian
is written as:
$$\eqalign{{\cal H}_I=&\lambda \int d^2 b\,h_b({\bf b})h_f({\bf b})\cr
  h_v({\bf b})=&c^{\dagger}_v({\bf b})a_v({\bf b})+
  a^{\dagger}_v({\bf b})c_v({\bf b})\;.}\eqno(2.3)$$
With this choice the interaction and the free Hamiltonian commute:
$[{\cal H}_o,{\cal H}_I]=0\;,$
the theory however is not trivial, even though it has been simplified, the 
S-matrix
can 
be written out in the form ${\cal S}=\exp[-\i {\cal H}\tau]$ where $\tau$ is an
 interaction time.
 
 \vskip 1pc
{\it 2.3 Discretization}
\vskip 1pc

The complete locality of the interaction in the transverse coordinates is both
unrealistic and sometimes inconvenient, we consider an alternative with
a finite size $\Delta $ of the hard interaction and a
discretization of the transverse plane. The size $\Delta $ is related to the
cut-off in the transverse momentum which must be put in order to be allowed to
perform perturbative calculations, so the natural choice is $\Delta \approx 
p_{\bot}^{-2}$; this choice leads also to the interpretation of
$\tau\approx 1/p_{\bot}\approx \sqrt {\Delta}$. The commutation relations 
become
 $\;[A_{v,j},A^{\dagger}_{u,i}]=\delta _{i,j} \delta _{u,v}\;,$ and so on.
 So in this discrete version the emission and absorption
 operators are dimensionless; also
 the coupling constant $g$ is dimensionless, it is related to the previous
 coupling constant by $\lambda=g\sqrt {\Delta}$. The parameter $\Omega$ plays 
no
 role in the next treatment, it might also coincide with the previously
 introduced $\omega$. In this way we get
$$\eqalign{{\cal H}_o=&\sum_{v,j} \Omega 
[A^{\dagger}_{v,j}A_{v,j}+C^{\dagger}_{v,j}C_{v,j}]\cr
          {\cal H}_I=&(g/\sqrt{\Delta})\sum_j H_{b,j}\cdot H_{f,j}\cr
          H_{v,j}=&C^{\dagger}_{v,j}A_{v,j}+A^{\dagger}_{v,j}C_{v,j}}
          \eqno(2.4)$$
Also the S-matrix becomes discretized and it takes the form
$${\cal S}=\prod_j {\cal S}_j\qquad ,\qquad 
{\cal S}_j=\exp[-\i (g/\sqrt{\Delta})\tau H_{b,j}\cdot H_{f,j}]\;,\eqno(2.5)$$
With the previous interpretation of $\tau$, one obtains the simpler expression:
$${\cal S}_j=\exp[-\i g H_{b,j}\cdot H_{f,j}]\;.\eqno(2.5')$$
In order to apply this model one must choose a definite initial state; it will
be factorized in the same way as the S-matrix: as far as its structure in a 
site
$j$ is concerned there are no strong indications. A possible choice, related to
 some theoretical ideas about the non perturbative partonic
 structure of the hadron[7,8], is the coherent state, so we may write
 $$|I>=\prod _j |I>_j\quad ,\quad |I>_j=\exp [-\um (|F_b|^2 +|F_f|^2)] 
       \exp[F_bA^{\dagger}_b +F_f A^{\dagger}_f]\,|>_j \eqno(2.6)$$
The vacuum state is also formally factorized; it has to be noted that the 
weight
 $F$ of the coherent state may vary from site to site. For simplicity
 the index $j$ will be not written out, whenever possible.
\par
It is useful to introduce the auxiliary operators through the definitions
$$\eqalign{ P=(C+A)/\sqrt{2}\quad &Q=(C-A)/\sqrt{2}\cr
   A=(P-Q)/\sqrt{2}\quad &C=(P+Q)/\sqrt{2}}\eqno(2.7)$$
   in this way we get
   $$ {\cal H}_o=\sum_{v,j} \Omega
   [P^{\dagger}_{v,j}P_{v,j}+Q^{\dagger}_{v,j}Q_{v,j}]\quad ,\quad
        H_{v,j}=P^{\dagger}_{v,j}P_{v,j}-Q^{\dagger}_{v,j}Q_{v,j}\eqno(2.8)$$
and it is also easy to express $|I>_j$ in term of the basis generated by $P$ 
and
   $Q$.
   
   \vskip 2pc
   {\bf 3. Inelastic cross section}
   \vskip 1pc
{\it 3.1 Contribution of a discrete site}
\vskip 1pc

The first observable that can be calculated within the model is the inelastic
cross section; it will be calculated using the discrete formulation for the 
states and for the S-matrix. In the basis generated by the operators $P$ and 
$Q$
the operator ${\cal S}_j$ is diagonal, so it is easy to calculate the matrix 
element $S_j={}_j\!<I|{\cal S}_j|I>_j$, it has the expression:
$$S_j=N^2 \sum _{k_1\cdots k_4}{1\over {k_1!k_2!k_3!k_4!}}
  (\um |F_b|^2)^{k_1+k_2} (\um |F_f|^2)^{k_3+k_4}\exp[-\i g(k_1-k_2)(k_3-k_4)]
  \;.\eqno(3.1)$$
 The indices $k_1,k_2,k_3,k_4$ refer to the quanta created respectively
by $ P_b,Q_b,P_f,Q_f$. The normalizing factor is given in eq.(2.6), actually
$$N=\exp \bigl[-\um (|F_b|^2+|F_f|^2)\bigr]\;.\eqno (3.1')$$
By using the following representation, with $\alpha \beta =g$, 
$$\exp[-\i g(k_1-k_2)(k_3-k_4)]=(2\pi )^{-1}\int dudv \exp \bigl[\i uv +
\i \alpha u(k_1-k_2) +\i \beta v (k_3-k_4)\bigr]$$
the multiple sum in the expression of $S_j$ can be transformed into an
integral. We use the positions $T_v=|F_v|^2$ and we obtain:
$$S_j=\udp\int dudv \exp [\i uv] 
 \exp \bigl[-T_b (1-\cos\alpha u)-T_f (1-\cos\beta v)\bigr]\;.\eqno(3.2)$$
\vskip 1pc
{\it 3.2 Continuum limit}
\vskip 1pc
When the distribution functions $F_v$ do not vary strongly from site to site 
one
can devise a continuum limit \footnote*{This means precisely that the
distributions are smooth, not that we consider the unrealistic limit $p_{\bot}
\to\infty$} so that the partonic structure of the colliding hadrons is 
described
as:
$$|I>=
\exp \int d^2b\Bigl[[-\um [|f_b({\bf b})|^2+|f_f({\bf b-B})|^2]+
[f_b({\bf b})a^{\dagger}_b({\bf b})+f_f({\bf b-B})a^{\dagger}_f({\bf b})]\Bigr]
 | >\;.$$
Here ${\bf B}$ denotes the relative impact parameter of the two hadrons. We
consider the natural relation $F\approx f \sqrt{\Delta}$; therefore, if $f$ is 
not singular, in the expression for $S_j$ the factors $T$ become small, so the
exponential in the integral representation can be expanded and integrated term
by term giving as a result: 
 $$S_j\approx 1-T_b T_f (1-\cos g)+\um T_b T_f
 (T_b+T_f)(1-\cos  g)^2+\dots\eqno(3.3)$$ 
 With our normalization the inelastic cross section at fixed
 hadronic impact parameter $B$ is 
  $$\sigma({\bf B})=2<I|(1-\Re {\cal S})|I>-\bigl | <I|(1-{\cal S})|I>\bigr |^2
  \;.\eqno(3.4)$$
  The product of the matrix elements ${\bf S}=\prod_j  S_j$ is, as usual, 
  calculated through
  the sum of the logarithms and the sum $\Delta\sum_j$ is finally
  converted into the integration $\int d^2b$. The final result is: 
  $$ \eqalign {\sigma ({\bf B})=1-\exp\int d^2b \Bigl[-&\hat \sigma t_b({\bf 
b})
  t_f({\bf b-B}) +\cr
  &\uq\hat \sigma ^2 t_b({\bf b}) t_f({\bf b-B}) 
  \bigl(t_b({\bf b}) + t_f({\bf b-B}) \bigr)+\cdots\Bigr] \;.}\eqno(3.5)$$
  Two definition are introduced, $\hat \sigma=2\Delta (1-\cos g)$, since this 
  is precisely the parameter which has the role of elementary partonic cross 
  section, and
  $t_v({\bf b})=|f_v({\bf b})|^2$, giving the transverse density of bound
  partons.
  \par
 The form of the inelastic cross section is quite usual. The second term in the
 argument of the exponential represents the rescattering corrections, which
 will be discussed below, where a nonuniform model of the hadron will
  be explored.
 The cross section arises from the integration over the impact parameter, 
 so the result depends to a large extent on the properties of the density 
functions
 $t_v$.
 \par
 As a simplest case the distribution can be taken to be completely uniform in
 $b$:
 $$t_b({\bf b})=\rho_b \vartheta(R-|b|)\quad,\quad
 t_f({\bf b})=\rho_f \vartheta(R-|b|)\;;\eqno(3.6)$$
 elementary geometrical considerations give $|B|=2R\cos\um\gamma$ with 
 $0\le\gamma\le\pi$. The exponent in the integrand is given by the 
 partial superposition of the two disks, the superposition area is
 $$W= R^2 (\gamma -\sin\gamma)\;.\eqno(3.7)$$ 
 and the cross section is expressed as:
 $$\sigma_{in}=2\pi R^2\int_o^{\pi}d\gamma \sin \gamma 
 \Bigl[1-\exp[-\nu \xi]\Bigr]\;,\eqno(3.8)$$
 where $\nu =\hat \sigma \rho_b \rho_f \pi R^2$ and $\xi=(\gamma
 -\sin\gamma)/\pi=W/(\pi R^2)$.
 \par
 With the previous interpretation of $\hat\sigma$, the numerical constant 
 $\nu\xi $
 is the mean number of partonic interactions. In the limit of wholly uniform 
 density the introduction of corrective terms like those appearing in the 
 expression of $S_j$, eq. (3.3), amounts simply to a redefinition of $\nu$.
 \par
 It is possible to give a simple analytical form for $\sigma_{in}$ in the two
 limiting situations of very small or very large $\nu$. In the first case one
 gets 
 $$\sigma_{in}=\pi R^2\cdot \nu \;.\eqno (3.8')$$ 
 In the second case we can start from
 the expression 
 $$\sigma_{in}=2\pi R^2[2-D(\nu)]\eqno (3.8'')$$ 
 where the real function 
 $D(\nu)$ defined by eq(3.8) is monotonically decreasing, for large $\nu$ it 
 results $D(\nu)\approx 2 (6\nu^2 /\pi^2)^{-1/3}\cdot\Gamma (\dte)$ so that 
the
 geometrical limit of black disks $4\pi R^2$ is approached.
 \vskip 2pc
 {\bf 4. Inclusive cross sections and multiplicity distribution}
 \vskip 1pc
 {\it 4.1 Pair and double-pair production}
 \vskip 1pc
 The production of a pair is signaled both by the production of a backward
 parton and by the production 
of a
 forward parton, the Hamiltonian being fully symmetric. We make then the
 arbitrary choice of looking at the forward particles only; successively we 
shall investigate
 how much the rescattering processes may destroy the sharp correlation
 between backward and forward scattered partons.
 We start from the computation of the inclusive production from a single site
 \footnote*{The production of a double pair in a single site is not included
 because the size $\Delta$ is defined by the hard interaction, so the 
production
 of four particle in a single site would not give rise, if not accidentally, to
 two distinct pairs, each with compensating momenta.}
 $$<X_j> ={}_j\!<I|{\cal S}_j^{\dagger} C_f^{\dagger}C_f{\cal S}_j|I>_j \eqno
 (4.1)$$
 and we observe that $H_b$ can be treated as a number with respect to the
 forward operators, so the following relation holds:
$${\cal S}_j^{\dagger} C_f{\cal S}_j= C_f \cos(g H_b)+\i A_f \sin(g H_b)\;.
\eqno (4.2)$$
The state $|I>_j$ is a coherent state of bound partons and contains no free 
parton so one obtains:
$$<X_j> ={}_j\!<I|A_f^{\dagger}A_f\sin^2(g H_b)|I>_j\;.\eqno (4.1')$$
The matrix element in the previous expression is better calculated in the basis
generated by the operators $P$ and $Q$ (details are given in the Appendix), 
and the result is
$$<X_j>
=\um T_f\bigl[1-\exp[-T_b(1-\cos (2g)]\bigr]\;. \eqno (4.3)$$
We can now go to the continuum limit, always under the hypothesis of smooth
distributions $t_v({\bf b})$; we find
a problem in the presence of the function
$\cos (2g)$ instead of $\cos g$, in fact the elementary partonic
interaction
enters in $<X_j>$ through the quantity
$\kappa=\um \Delta [(1-\cos (2g)]$, which is related to $\hat \sigma$ in this 
way:
 $\kappa=\hat \sigma\cdot \um [1+\cos g]$. The two constants coincide
 for small $g$, where both take the same value: $\hat \sigma =\kappa= 
g^2\Delta$, higher
 powers in $g^2$ make however them different. Unitarity corrections at the 
level
 of parton-parton collisions are in fact different in the total and in the
 inclusive cross section.
 \par
 In the simple case of small values of $T$ we expand the exponential of eq. 
(4.3)
  and we get the usual expression:
$$X({\bf B})=\kappa \int d^2b t_f({\bf b}) t_b({\bf b-B}) \eqno (4.4)$$
and when we consider the integration over the hadronic impact parameter 
$$D_1=\int X({\bf B}) d^2B \eqno (4.5)$$ 
 the two integrals get factorized.
When, however, in the same conditions we calculate the double pair production 
we
do not end with a factorized expression (there are four factors and three
integrations). In this case we obtain
$$D_2=\kappa^2\int d^2B d^2b d^2b'
t_f ({\bf b}) t_b ({\bf b-B})t_f ({\bf b}') t_b ({\bf b'-B});.
 \eqno (4.6)$$
 In the limit of rigid disk, using the geometrical considerations and the 
 definition of the previous section 3.2 the expression becomes:
 $$D_2=2 (\pi R^2)^3 (\rho_b \rho_f\kappa^2)^2\int \xi^2 \sin\gamma d\gamma =
 (\pi R^2)^3 (\rho_b \rho_f\kappa)^2 [1-16/(3\pi^2)] \;.\eqno (4.7)$$
A ratio of the quantities that have been now calculated and that is of
phenomenological interest is $\sigma_{\rm eff}=[D_1]^2/D_2$, [9,10] which has
the nice property, within this treatment, of being independent of $\kappa$.
It 
depends rather on the space behavior of $t({\bf b})$, namely on the hadron
shape.
Since in the rigid disk limit it results from eq (4.4,5) that
 $$D_1=(\pi R^2) \kappa \rho_f \rho_b \eqno (4.5')$$
  the ratio we are looking for is given by:
$$\sigma_{\rm eff}={{\pi R^2}\over {1-16/(3\pi^2)}}
\approx 2.2\pi R^2\;,\eqno(4.8)$$

The previously calculated expression of $\sigma_{in}$ is really the
hard part of the total inelastic cross section, where as "hard" part we mean
the contribution of all the events with at least one hard scattering.
If we believe that, in going on with the total energy these events become
dominating we would like to have this term not too small with respect to the
experimental $\sigma_{in}$ ,which in turn appears to be sizably larger, of 
about
a factor 2, with respect to $\sigma_{\rm eff}$, so in this model we expect
to approach at high energy the black-hadron limit which produces 
$\sigma_{in}=4\pi R^2$. 
\vskip 2pc
{\it 4.2 Multiplicity distribution}
\vskip 1pc
The distribution of the multiplicities of the produced pairs is calculated
by defining the projection operator over the number of free partons. Since 
there is a sharp distinction between backward and forward particles we can
choose to take the forward parton as a signal of the pair production.
For a fixed site $j$ the number projector is
\footnote*{Whenever possible the indices $f,b,j$ are suppressed}
$${\cal P}_n={1\over {n!}}:C^{\dagger}C\e^{-C^{\dagger} C}:\eqno (4.9)$$
The colon indicates the normal ordering of the $C$-operators, more precisely
if the operators refer to the forward particles this is the forward normal 
ordering,
and in no way it affects the backward operators. 
The properties of the ${\cal P}_n$ operators are easily verified. They are
evidently diagonal in the number basis and clearly $<m|{\cal P}_n |m>=0$ when
$m<n$, for $n\le m=n+\ell  $ we get, through direct computation 
$$<m|{\cal P}_n |m>={1\over {n!}}\sum_{k=0}^{\ell } (-)^k
{{(n+\ell)!}\over{(\ell-k)! k!}}={{(n+\ell)!}\over{\ell! n!}}[1-1]^{\ell}=
\delta_{\ell ,0}\;.$$
It could be more convenient to deal with the generator of the projectors:
$${\cal P}_n={1\over {n!}}\Bigl({\partial \over {\partial \mu}}\Bigr)^n
{\cal Z}\Bigr|_{\mu=-1}
\quad {\rm with}\quad {\cal Z}=:\e^{\mu C^{\dagger}C}:\;.\eqno (4.10)$$
An auxiliary function is introduced
$$Z (\mu )= <I|{\cal S}^{\dagger} {\cal Z\,S}|I>\eqno(4.11)$$

Calculations are strongly simplified by the normal ordering; 
through the same steps which lead from eq.(4.1) to eq.(4.3)
 one can get in fact
$$Z (\mu )= \exp[\um \mu T_f] <I|\exp[-\um \mu T_f \cos(2g H_b)]|I> 
\;\eqno(4.12)$$
and one must remember that the functions of $H_b$ are {\it not} normal ordered.
The matrix element of eq (4.12) is computed by expanding the operator into 
numerical Bessel functions and operatorial trigonometric functions, using the 
relation[11]:
$\e^{z\cos \theta}=I_o(z)+2\sum_k I_k(z) \cos (2k\theta)$, which allows to 
perform the same calculations leading from eq.(4.1) to eq.(4.3). The
final expression is written as
$$\eqalign{ Z (\mu )= &\exp[\um \mu T_f]\Bigl[I_o(-\um \mu T_f)\cr
+&2 \sum_k I_k(-\um \mu T_f) \exp[-T_b (1-\cos(2kg))]\Bigr]}\;.
\eqno(4.12')$$
According to eq (4.10) the derived multiplicity distribution, as seen from a
single site is
$$K_n={1\over {n!}}\Bigl({\partial \over {\partial \mu}}\Bigr)^n 
Z (\mu)\Bigr|_{\mu=-1}\;.$$
When the production at the single site in not very strong it is natural to
expand in the absorbing term $T_b$. A straightforward although a bit lengthy
calculation yields, to the second order in $T_b$:
$$\eqalign{ K_n=&{1\over {n!}}[\um T_f]^n \Bigl[T_b(1-T_b)
[1-\cos(2g)]^n \exp\Bigl(-\um T_f[1-\cos(2g)]\Bigr)\cr
+&\uq T_b^2 [1-\cos(4g)]^n 
\exp\Bigl(-\um T_f[1-\cos(4g)]\Bigr)\Bigr]+\cdots}\eqno(4.13)$$
The result is evidently a sum of two Poissonian distributions.
The origin is the Poissonian distribution of the initial coherent state,
which is modified in a well defined way by re-interactions.
\par 
When re-interactions are important there is no simple expression for the 
result,
as usually we get a
not too awkward expression in the extreme limit ,$i.e.$, very large $T_b$.
In this case we could neglect in eq.(4.12') all the terms containing the
negative exponent of $T_b$. We translate[11] afterwards the expression 
containing the Bessel function into one expressed through the confluent 
hypergeometric function:
$$Z (\mu)\approx M(\um;1;\mu T_f)\;,$$
and we obtain in this way for the multiplicity distribution the form:
$$K_n\approx {1\over {n!}} {T_f}^n\exp[-T_f]\cdot L_n\quad ,\quad 
L_n={1\over {n!}}\bigl(\um \bigr)_n M(\um ;n+1; T_f)\eqno (4.14)\;.$$
The expression has been put into a form of a Poissonian distribution times
another factor, this further factor factor $L_n$ is not a small correction
however, 
it changes in an essential way the shape of the distribution.
\vskip 2pc
{\it 4.3 Forward-backward correlations}
\vskip 1pc
Until now pair production has been described by looking at the production
of a definite component of the pair. When the re-interactions are not very 
important this attitude is justified. If however
there is a strong localized production in one site further investigations
are needed. In this last case we can calculate within the model the variance 
and
the covariance of the number of emitted partons. 
We start from the computation of the dispersion in the inclusive production in
 a single site $j$:
 $$<X^2> =<I|{\cal S}^{\dagger} C_f^{\dagger}C_f C_f^{\dagger}C_f{\cal S}|I>\;.
  \eqno (4.15)$$
  The calculation goes along the same patterns as in the previous 
  calculation of $<~X~>$. The result is 
 $$<X^2> =T_f^2<I|\sin^4(g H_b)|I>+T_f<I|\sin^2(g H_b)|I>$$
 and for the variance one obtains:
 $$\eqalign{\Sigma _f=&<X^2>-<X>^2
=\um T_f\bigl[1-\exp[-T_b(1-\cos (2g)]\bigr]+\uo T_f^2 \times\cr 
\Bigl(&1-2\exp\bigl[-2T_f[1-\cos(2g)]\bigr]+
 \exp\bigl[-T_f[1-\cos(4g)]\bigr]\Bigr)
\;.} \eqno (4.16)$$
The starting point to calculate the covariance is
 $$<W> =<I|{\cal S}^{\dagger} C_f^{\dagger}C_f C_b^{\dagger}C_b{\cal S}|I>\;,
  \eqno (4.17)$$
  Because of the coherent state structure of $|I>$ one can show that
 $${\cal S}^{\dagger} C_b {\cal S}|I>=\i F_b \sin (g H_f)|I>\;. $$
 By using this relation the expression of $<W>$ gets simplified to some extent.
 With some more work one recognizes also that
$$\eqalign {A_f \sin (g H_f)|I>=&F_f \cos g \sin (g H_f)|I>\cr
            C_f \sin (g H_f)|I>=&F_f \sin g \cos (g H_f)|I>}$$
 From now on the calculations uses the results already seen, like eq 
.(4.1,1',3)
 and  yields for the covariance $\Sigma _{f,b}=<W>-<X_f><X_b>$ the following 
 expression:
 $$\Sigma _{f,b}=\um T_f T_b \sin ^2 g \Bigl[\exp[-T_b\{1-\cos (2g)\}] +
 \exp[-T_f\{1-\cos(2g)\}]\Bigr] \eqno (4.18)$$
 In term of the quantities $\Sigma _{f,b}$ one can define the correlation 
coefficient as:
 $$\rho _{f,b}={{\Sigma _{f,b}}\over{[\Sigma _f \Sigma _b]^{1/2}}} \eqno 
(4.19)$$
 A look to eq (4.16,18,19) shows, as expected, that for small values of $T_v$,
 namely for small production, the correlation goes to 1, on the contrary when
 $T_v$ becomes large the correlation goes to zero.
 
\vskip 2pc
{\bf 5. Non uniform hadrons}
\vskip 2pc
 {\it 5.1 Inelastic cross section}
 \vskip 1pc
 We wish now to explore the possibility that the hadron, and its projection
 over the transverse plane, shows strong inhomogeneities in the matter density.
 This is represented by assuming the existence of black spots, that cover a
 limited amount of the transverse area, while a much fainter "gray" background
 smoothly fills the rest of the hadron. For the black spots the continuous 
 limit (sect. 3.2) is not justified, but since
 they cover globally a small area, we can distinguish three possible kinds
 of collisions:
 The spot-spot collision, to be treated individually, the spot-background
 collision, where the background is treated as a continuum in $b$ and finally
 the background-background which is nothing but the continuous limit already
 studied.
 In order to deal with the first case we start again from eq. (3.2), in the
 limit of very large $T_j$ the integral will be expanded around the 
 zeros of the exponent by setting:
 $$u\alpha=2\pi n+\chi \quad ,\quad v\beta=2\pi m+\phi \;;$$
 the subsequent Gaussian integrations over $\chi $ and $\phi$ give:
 $$S_j\approx {1\over g} {1\over {\sqrt {T_f T_b}}}
 \sum_{m,n}\exp [\i (2\pi)^2 mn/ g]
 \eqno (5.1)$$
 In order to give an estimate of its value, the double sum is then 
 converted into a double integration from $-\infty$ to $+\infty$
 with the final result:
 $$S_j\approx S_s={1\over{2\pi }}{1\over {\sqrt {T_f T_b}}} 
 \eqno (5.2)$$
 In the mixed spot-background collision, always taking eq (3.2) as starting
 point, it is possible to expand in one of the $T_v$ terms, keeping the full
 expression for the other one with the results:
  $$\eqalign{S_j\approx &S_f=1-T_b\bigl(1-\exp[-T_f(1-\cos g)]\bigr)
  \qquad {\rm or}\cr
S_j\approx &S_b=1-T_f\bigl(1-\exp[-T_b(1-\cos g)]\bigr)}\eqno (5.3)$$
  for the collision between a forward spot with the 
  backward background and between a backward spot and the forward background.
  The background-background contribution has been already given in eq.(3.3).
  \par
  We are now in position to calculate the modification to the expression for 
the
  inelastic cross section, eq.(3.8), that are introduced by the hypothesis of 
the
  existence of the black spots. The gray distribution is again assumed to be 
uniform
  so the term in eq.(3.3) is constant and it will be denoted by $S_o$.
  The interaction area $W$ is considered as composed by $w$ small elements
 $W=w\Delta$, the ratio $\xi=W/\pi R^2$, will also be frequently used, so that 
 $w=\xi (\pi R^2/ \Delta)$. 

The quantum mechanical treatment of a
nonuniform hadron is now presented in a form which contains certainly some 
rough simplifications that were unavoidable in order to deal with a system with
an elevated degree of complexity.
We may begin by
assuming, at fixed $j$, a state of the kind:
$|I>_j=|I_b>_j|I_f>_j$
$$|I_b>_j=\Bigl[x\exp [-\um |F_b|^2 ] 
       \exp[F_bA^{\dagger}_b]+y\exp [-\um |G_b|^2 ] 
       \exp[G_bA^{\dagger}_b]\Bigr]|> \eqno (5.4)$$
 In this expression, and in the similar one for $|I_f>$, the terms
 $F=|F|\e^{\i\phi}$ and $G=|G|\e^{\i\chi}$ denote two different thickness 
while
 the coefficients of the q.m. superposition are $x,y$. Since the two coherent
 states are not orthogonal the general form of the normalization condition is 
 complicated:
 $$\eqalign{1&=|x|^2+|y|^2+\cr
   &\exp \bigl[-\um(|F|-|G|)^2\bigr] 
 \bigl\{xy^{*}\exp\bigl[|FG|[1-\e^{\i(\phi-\chi)}]\bigr] 
 +x^{*}y\exp\bigl[|FG|[1-\e^{\i(\chi -\phi)}]\bigr]\bigr\}}$$
 When however the two thickness are very different the last term in the
 normalization condition is exponentially depressed and we are left with
$$|x|^2+|y|^2\approx 1\;.$$
It is reasonable to expect that the same feature occurs also in calculating the
relevant matrix elements, but in a significant case the calculations will be
carried out explicitly so that the guess can be verified. In fact a non 
diagonal
term of the matrix
element of ${\cal S}_j$ is computed along the same lines yielding eq.(3.1,2)
and the result is:
$$\eqalign {S_{ND}=&(2\pi )^{-1}\int dudv \exp [\i uv] 
 \exp \bigl[-\um (|F_b|-|G_b|)^2-\um (|F_f|-|G_f|)^2 \bigr] \times\cr
 &\exp \bigl[-|F_b G_b|\bigl(1-\e^{\i(\phi -\chi)_b}\cos\alpha u\bigr)\bigr]
  \exp \bigl[-|F_f G_f|\bigl(1-\e^{\i(\phi -\chi)_f}\cos\beta v\bigr)\bigr]}
 \eqno(5.5)$$
 It is therefore enough to have one $F$ very different from one $G$ in order to
 obtain an exponentially suppressed contribution.
 \par
 One can see in this way that out of the 16 terms, that are produced in 
calculating the
 complete matrix element of ${\cal S}_j$, 12 $i.e.$ the interference
 terms, are exponentially suppressed.
 \par
 It is useful to formalize this approximate result in the following way:
 let be $|I_b>_j=[\B +\mu \S]_b $ where 
 $$ \BB \S \approx \SS \B \approx 0 \quad {\rm and}\quad \BB \B \approx \SS \S
 \approx 1 \eqno(5.6)$$
 
Then ${}_j<I_b|I_b>_j \approx 1+\mu^2 $ and
the S-matrix element takes the form\footnote*
{The parameter $\mu$ measures the ratio of the area covered by the spots to the
background; it is assumed small and real because the relative phases have no
role}
 $$S_j=[\BB|+\mu\SS|]_{b,j}[\BB|+\mu\SS|]_{f,j}{\cal S}_j 
 [\B +\mu\S]_{b,j}[\B +\mu\S]_{f,j}(1+\mu^2)^{-2}$$
 which more explicitly gives
 $$S_j=[S_o+\mu^2 S_b +\mu^2 S_f +\mu^4 S_s](1+\mu^2)^{-2}\;.\eqno(5.7)$$

In the limit of totally black spots, where we have the following
 values for the terms entering in eq.(5.7), see eq.s (3.3,3.6,5.3,5.2)
 $$ \eqalign {S_o=&1-\hat\sigma \rho_b \rho_f \Delta \cr
      S_b =&1-\rho_f\Delta =S_o \eta_b\cr
      S_f =&1-\rho_b\Delta =S_o \eta_f\cr
      S_s=&0 } \eqno (5.8)$$

  the expression for the S-matrix is:
  $$ \eqalign {{\bf S}=&\prod_j S_j=S_o^w\cdot [1+\mu^2 \eta_b+\mu^2
  \eta_f]^w(1+\mu^2)^{-2w}\cr
  =&{\bf S_o}{\bf S_c}}\;. \eqno (5.9)$$
  where the factor ${\bf S_o}\equiv{S_o}^w$ gives the contribution of the 
  background scattering.
  \par
  A much simpler expression can be written when $\mu$ is small and keeping into
  account that ${\bf S_o}$ is not very different from 1:
   $${\bf S_c}=1-w \mu^2 \Delta
  [\rho_b(1-\hat\sigma \rho_f)+ \rho_f(1-\hat\sigma
   \rho_b)] \eqno (5.10)$$
   One may notice that the first factor, $w \mu^2 \Delta$, represents the part
   covered by black spots within the interacting area of the hadrons at given
   {\bf B}. A more
   general observation is that the systematic use of the relations of 
eq.(5.5,6)
   eliminates the more complicated aspects induced by quantum mechanics and the
   answer is the same as if one would have taken a probabilistic distribution 
of
   spots in the transverse section of the interacting hadrons.
   
   \vskip 2pc
 {\it 5.2 Pair and double pair production}
 \vskip 1pc
 The basic ingredient for these new calculations is always given by the
 expression in eq. (4.3), which must be particularized for the situations under
 examinations. For simplicity and, more, in order to put in evidence the 
 features of this particular model the background will be taken as thin so that
 in eq. (4.3) the first term of the expansion of the exponential is enough
 while in the case of the spots the exponential will be considered totally
 absorbing. The local production amplitude is the sum of 4 terms, it will be
 indicated as 
 $$<X_j>=[X_{o,o}+\mu^2 X_{s,o}+\mu^2 X_{o,s}+\mu^4 X_{s,s}](1+\mu^2)^{-2}
 \eqno (5.11)\;.$$
 The first term represents the pure background interaction, it is given by
 eq.(4.4), the second term
 represents a forward spot interacting with the backward background so $T_f$ is
 large while $T_b=t_b\Delta$, the third term represents the opposite 
situations,
 note that the resulting expression is $not$ symmetrical because we are looking
 to the forward produced particles, finally the fourth term gives the effect 
of
the spot-spot interaction. The four terms must be summed over the allowed 
values
 of $j$, $i.e.$ over the position included in the interaction region. The sums
 may be expressed as :
 $$\eqalign {\sum_j X_{o,o}=&\kappa \rho_f \rho_b W\cr
            \sum_j X_{s,o}=&\kappa \rho_b (T_f/\Delta) W\cr
            \sum_j X_{o,s}=&\um \rho_f W\cr
            \sum_j X_{s,s}=&\um (T_f/\Delta) W\;.} \eqno (5.12)$$ 
            The dependence on the angular variable $\gamma$ is the same for
            the four terms, $i.e.$ we can in any case write
            $\sum X= K\xi$.
 The integration over the impact parameter is factorized, so we get for the 
 inclusive production rates expressions as in eq.s (4.5',7). Actually 
 $$D_1=\pi R^2 K\quad,\quad D_2=\pi R^2 K^2 [1-16/(3\pi^2)] $$
 The result for $\sigma_{\rm eff}$ is therefore precisely the same as in 
eq.
  (4.8). It has to be noticed that although $T_f$ may be large $(T_f/\Delta)$ 
in
  finite and would remain finite even in a formal limit $\Delta\to 0$.
  At first sight it seems that nothing is gained by introducing an
  inhomogeneity into the hadron, but in fact some news features are present.
  The expression of $\sigma_{\rm eff}$ is purely geometrical, it does not
  contains $\mu$, $R$ is simply the
  radius of the area where the spots may be found; the expression of 
  $\sigma_{in}$ on the contrary contains dynamical parameters. It is 
  therefore instructive to compare $\sigma_{in}$ and $\sigma_{eff}$.
  \par
  A clear, although unrealistic example is the limiting case in which 
  the background is so thin that it
  contributes negligibly to the inelastic cross section, then the S-matrix
  element, depends only on the spot-spot interaction and it has the form
  $${\bf S}=\Bigl[1-{{\mu^4 (1-S_s)}\over {(1+\mu^2)^2}}\Bigr]^w
  \eqno (5.13)$$ 
  For small values of $\mu$ this expression may approach 1 and so the inelastic
  cross section becomes small, this situation would correspond to have few 
spots
  wholly black distributed in a wide and very thin background.
   The conclusion of this analysis is then that in order to have
  $\sigma_{\rm eff}<\sigma_{in}$ the hadron should be compact, $i.e.$ without
  holes or transparent regions. There is however
  another possibility. The spots can in fact be distributed with some 
correlation
  among themselves, this possibility can be investigated, to some extent, 
within
  the model. The analysis is however formally awkward and not very
  conclusive, so it will be just mentioned but not reported explicitly.
  
  \vskip 2pc
  
  {\it 5.3 Local effects of unitarity}
  \vskip 1pc
  Until now the models of the hadron that have been considered in more detail 
  cover two extreme situations, the case where there is not local rescattering
  and unitarity is relevant only to the whole hadronic interaction,
  allowing a description of the multiple disconnected partonic collisions, and
  the case where the re-interaction is so strong that, locally, the hadron is
  completely absorbing. We think useful to investigate briefly some 
intermediate
  situation, in this context sometimes the matter distribution in the hadron
  (the bound-parton distribution of the model)
  will be assumed to have a Gaussian shape, just in order to allow some 
  explicit analytic
  calculation.
  \par 
  We start from the inclusive productions: for the single inclusive we have in
  general in the model, comparing eq.s (4.3,4,5)
  $$ D_1=\um  \int t_f ({\bf b+B}) \Bigl[1-
         \exp [-2\kappa t_b ({\bf b})]\Bigr]d^2 b d^2 B\eqno (5.14)$$
         and for the double inclusive it results
  $$ D_2=\uq \int  t_f ({\bf b+B}) t_f ({\bf b'+B})
  \Bigl[1-\exp [-2\kappa t_b ({\bf b})]\Bigr] 
  \Bigl[1-\exp [-2\kappa t_b ({\bf b'})]\Bigr]d^2 b d^2 b' d^2 B\eqno (5.15)$$
  The densities $t$ vary with the total energy of the process because, really,
  only the part of the partonic spectrum  that can give rise to the hard
  scattering enters in eq.s(5.14,15) and this part certainly grows with the
  total energy of the collision. The simplest way in which this variation can 
be
  implemented is by rescaling the densities by a factor $\lambda$ growing with
  the energy. It is evident that in calculating $\sigma_{\rm eff}=[D_1]^2/D_2$ 
  the factors affecting $t_f$ are eliminated in the ratio  and the overall 
  effect of $\lambda$ amounts formally to a rescaling of $\kappa$.
  \par
  So, until $2\kappa t ({\bf b})<<1 $ one expands the exponential and gets the
  expression already displayed which contains only geometrical elements, but 
  when $2\kappa t ({\bf b}) \approx 1 $ the final expression is less simple.
  It will be studied with the particular choice $ t ({\bf b})=(\mu/\pi)
  \exp [-\mu{\bf b}^2]$ and with the definitions $ {\bf b=R+r}/2 \quad 
  {\bf b'=R-r}/2$, so that 
  $$\int  t_f ({\bf b+B}) t_f ({\bf b'+B}) d^2 B=(\mu/2\pi )\exp [-\um \mu{\bf
  r}^2]= {\cal F}({\bf r})$$
  and       
   $$ D_2=\uq \int {\cal F}({\bf r}) 
  \Bigl[1-\exp [-2\kappa t_b ({\bf R+r}/2)]\Bigr] 
  \Bigl[1-\exp [-2\kappa t_b' ({\bf R-r}/2)]\Bigr]d^2 R d^2 r$$
  When $2\kappa t ({\bf b})<\tau $ and $\tau$ is not too small, the
  integral can be separated into two parts, the first is approximately 
  $$\int {\cal F}({\bf r}) [1-\e^{-\tau}]^2 d^2 R d^2 r$$
  with the bounds $b<b_o \;,\; b'<b_o\;,\; 
  b_o^2=\ln(2\mu\kappa/\pi \tau)/\mu$.
  The integration region in $d^2 R$ has the same geometrical shape as the
  integration region yielding eq (3.7), only the role of the radius is played 
by
  $b_o$. The integration in $d^2 r$ is always convergent and has no significant
  dependence on $b_o$, so the conclusion is that this first addendum is
  proportional to $b_o^2$ and so to $\ln\kappa$. The second addendum is
  certainly not growing with $\kappa$, at fixed $\tau$, so at the end we get 
the
  result $D_2\propto b_o^2\propto \ln\kappa$.
  The behavior of $D_1$ is much simpler to estimate, the double integral is
  factorized, the first factor does not depend on $\kappa$, the second is
  proportional to $\ln\kappa$, for large $\kappa$ as it may be seen by explicit
  calculation.
  The conclusion of the analysis shows that the ratio giving $\sigma_{\rm 
eff}$ has
  a region where it stays essentially constant also if the hadron 
  has no sharp boundaries but that at the end it will suffer a logarithmic
  increase whose origin has to be found in the local ($i.e.$ at fixed ${\bf 
b}$)
  unitarity. The moment at which these local unitarity effects begin to be
  relevant is when $2\kappa t ({\bf b})$ is not too small with respect to 1. 
  It must be remarked that the growth with energy is basically different for 
  $\sigma_{in}$ and for
  $\sigma_{\rm eff}$ since in the first case a logarithmic expansion is given
  already by the simple formula, which is "unitary" at hadronic level but not
  yet at partonic level, whereas in the second case the partonic unitarity is
  essential.
  The logarithmic growth is due to the particular choice of a Gaussian
  shape, an exponential shape is required to give rise to the 
square-logarithmic
  growth.
   
  \vskip 2pc
  {\bf 6. On the longitudinal dynamics }
   \vskip 1pc
  In the frame of the eikonal representation that has been used throughout the
  paper the longitudinal variables for the partons are in general expressed by
  means of the fraction of the longitudinal momentum carried by the partons.
  This is the natural choice for the bound partons, for a free parton the 
  usual variable is the rapidity whose connection with the fractional momentum
  requires the introduction of the transverse motion. Since we are always
  interested in situations where the partons are tied in a clear way to the
  initial hadrons we shall use the fractional momentum $x$ in every case,
  with the sharp distinction in forward and backward partons.\footnote*
  {With this limitations it is not possible to introduce the hard production
  into the parton dynamics. It is possible to take into account 
re-interactions
  among the produced partons; in so doing the S-matrix element acquires a
  further phase.}
  
  Starting from the formulation which is discrete in the transverse variables 
it
   is possible to introduce operators which depend on the fractional momentum 
   and give the usual commutation relations like
  $\;[A_{v,j}(x),A^{\dagger}_{u,i}(x')]=\delta
  _{i,j} \delta _{u,v} \delta (x-x')\;,$ and so on.
 and write the new forms of the Hamiltonians:
$$\eqalign{{\cal H}_o=&\sum_{v,j} \int dx\Pi_v x
[A^{\dagger}_{v,j}(x)A_{v,j}(x)+C^{\dagger}_{v,j}(x)C_{v,j}(x)]\cr
          {\cal H}_I=&(g/\sqrt{\Delta})\sum_j H_{b,j}\cdot H_{f,j}\cr
          H_{v,j}=&\int dx \bigl[C^{\dagger}_{v,j}(x)A_{v,j}(x)+
          A^{\dagger}_{v,j}(x)C_{v,j}(x) \bigr]}\;.\eqno(6.1)$$
  The hadron momenta are denoted by $\Pi_v$ and the fractions $x$ refer to the
  forward and backward total momenta according to the operator where they
  appear.
  In the same way the incoming states are described in terms of some
  distributions of longitudinal partons, the direct generalization of eq. 
(2.6),
  where the transverse factorization is maintained for the hadronic state 
  $|I>=\prod_j |I>_j$ and for the vacuum:
 $$|I>_j=\exp\Bigl[-\um \int dx(|F_b(x)|^2 +|F_f(x)|^2)]+\int dx 
 [F_b(x)A^{\dagger}_b(x) +F_f(x) A^{\dagger}_f(x)]\Bigr]\,|>_j 
 \;.
 \eqno(6.2)$$
 With this particular choice of the interaction term everything proceeds as
 before because it is again possible to define the auxiliary operators $P(x)\,,
 \,Q(x)$. If we exclude black spots and go to the continuum limit the answer 
is:
  $$ \sigma ({\bf B})=1-\exp\bigl[-\hat \sigma \int d^2b dx\, t_b({\bf b},x) 
  t_f({\bf b-B},x) 
  +\um\hat \sigma ^2\cdots\bigr]\;. \eqno(6.3)$$
  At this point, having put into the game the longitudinal variables it is 
clear
  that the choice $t_v=const$ finds no justification whatsoever; the most
  uniform choice corresponds to a uniform sphere, 
  $i.e.\;t_v({\bf b},x)=\rho_v(x)\sqrt{R^2-b^2}\vartheta (R-|b|)$
  The role played before by the partial superposition of two disks, described
  by $W=\pi R^2 \xi$ is now played by the product of the two volumes that are
  superimposed 
  $$U=\int\sqrt {R^2-{\bf b}^2}\sqrt {R^2-({\bf B-b})^2}\vartheta (R-|b|)
  \vartheta (R-|B-b|) d^2b\;. \eqno (6.4)$$
  This term can be written in a variety of alternative forms, but not 
completely
  by means of usual functions, a representation that turns out to be useful is:
  $$U=R^4\int \e^{\i\lambda\cdot {\bf B}}\bigl[j_1(\lambda R)/\lambda \bigr]^2 
  d^2\lambda =2\pi R^4\int J_o(\lambda B) j_1(\lambda R)^2
  d\lambda/\lambda\;.\eqno (6.4')$$
  The spherical Bessel function has the form
  $j_1(x)=x^{-2}\sin x-x^{-1}\cos x$. 
  The term $U$ appears in the definition of the inelastic cross section and
  gives rise to a very cumbersome expression ,a much simpler form is found 
  in the definition of the inclusive production and of the double inclusive
  production.
  The first term can be obtained also by direct integration:
  $$D_1=(4\pi R^3 /3)^2\kappa \int \rho_b(x)dx \int \rho_f(x)dx\;.\eqno (6.5)$$
  For the double pair production, the expression is:
  $$D_2=(4\pi R^3)^3 2R \Bigl[\kappa \int \rho_b(x)dx \int \rho_f(x)dx\Bigr]^2 
  \int_o^{\infty}\bigl[ j_1(t)\bigr]^4 dt/t^3\;.\eqno (6.6)$$
  The last integral gives:
  $${\cal K}=\int_o^{\infty}[ j_1(t)]^4 dt/t^3=0.01433$$
  So we finally obtain:

  $$\sigma_{\rm eff}={{2\pi R^2}\over {81 {\cal K}}}={2\pi R^2}/1.16\eqno 
(6.7)$$
  
    \vskip 2pc
  {\bf 7. Conclusions}
  \vskip 1pc
  
  The model presented and worked out in some detail allows a systematization of
  different aspects of the hard processes in multiparticle production and
  suggests also some interpretations in terms of hadron structure.
  The connection with QCD is not direct as it appears from the fact that the
  interaction term is quartic while the fundamental QCD interaction term is
  cubic. The reason is that in the actual approach the fundamental input is 
the parton hard
  collision, not the branching process.
  \par
  All unitarity corrections for the different processes are fully explicit,
  even
  though the final analytical expressions are given for the two extreme
  situations: weak perturbative corrections, very strong absorption. The
  analysis of the effects of inhomogeneities has been carried out purposely
  without specifying the possible origin of this supposed property, in fact
  the quantities which have been considered depend on a global characteristic, 
the total amount of
  black area, and not on other details. 
  The density distribution in the hadron is however relevant when 
  $\sigma_{\rm in}$ and $\sigma_{\rm eff}$ are compared.
  The representation of the hadron transverse area as a collection of elements
  of finite but much smaller extension requires a limitation of the momentum
  transfer which cannot go below, say, of 5GeV so that the interaction area 
  turn out of the order of (0.04 ${\rm fm)^2}$, small enough in comparison 
with
  the hadron extension.
  The assumption that the partons of the hadron build up, locally in ${\bf b}$,
  a coherent state is a theoretical prejudice. The test is is difficult 
  since the production process generally alters the
  multiplicity distribution, so that what finally one sees only
  a combination of
  hard and soft dynamics. To gain some better insight into the problem we have
  discussed, at the end of the $4th$ paragraph, the modification to the
  multiplicity distribution induced by hard rescatterings,
  \par
  Further questions can be raised: one is the possible coherence or
  correlation at different values of ${\bf b}$,what could, at the end, also 
  involve the role of the colour variables: in fact the transverse size of
  the hadron fraction over which the matrix element of ${\cal S}$ has been
  calculated is determined by the the transverse momentum transfer, which is a 
  quantity relevant for the perturbative treatment. Larger sizes of 
  transverse coherence may however show up in the
  multiple-production processes. Another question is why the incoming states
  should be locally coherent. More general initial states could be considered,
  but the new parameters one introduces in this way are at present out of
  control.
   \vskip 2pc
 {\bf Acknowledgments}
  \vskip 1pc
 This work has been partially supported by the Italian Ministry of University
 and of Scientific and Technological Research by means of the {\it Fondi per la
 Ricerca scientifica - Universit\`a di Trieste }.
 \vskip 2pc
\centerline{\bf Appendix }
\vskip 1pc
In this appendix we give some details on the calculation leading from eq (4.2)
to eq (4.3) since this result is repeatedly used and it yields also a model for
other similar calculations.
Since the calculation is performed at fixed site, the index $j$ will be 
omitted.
>From eq (4.1'), letting the operators $A_f$ act on the 
states $|I_f>$, it results furthermore.

$$<X_j> =T_f\,<I_b|\sin^2(g H_b)|I_b>\;.\eqno (A.1)$$

Going to the basis generated by the $P,Q$ operators it results also
$$<I_b|\sin^2(g H_b)|I_b>=\um-\uq\bigl(
<I_b|\exp[2\i g(P^{\dagger}_bP_b-Q^{\dagger}_bQ_b)]|I_b>+c.c.\bigr)\;.$$
The state is now expanded
$$|I_b>=N\sum_{m,n} {1\over{2^{(m+n)/2} m! n!}}F^{m+n}_b(P^{\dagger}_b)^m
(-Q^{\dagger}_b)^n|\;>$$
and the exponential acts now trivially on the Fock states so the final outcome 
is
$$<I_b|\sin^2(g H_b)|I_b>=\um-\uq N^2\sum_{m,n} {1\over{2^{(m+n)/2} m!
n!}}|F_b^2|^{m+n}\exp [2\i g (m-n)]+c.c.\;.\eqno (A.2)$$
A convenient way of computing the double summation of eq (A.2) is to perform
first the finite sum at fixed $l=m+n$, and then the infinite sum over $l$.
$$\eqalign{ <&I_b|\sin^2(g H_b)|I_b>=\cr
  &\um\Bigl[1-N^2\sum_l {1\over {2^l l!}}|F_b|^{2l}
\sum_{n=0}^l{l\choose n}\exp [2\i g (l-n)]\exp [-2\i g n]\Bigr]=\cr
&\um\bigl(1-\exp[|F_b^2|(1-\cos(2g))]\bigr)\;.}$$
The last step requires the insertion of the actual value of the normalizing
factor $N$, as it is given in eq (3.1'); then using eq (A.1) and the definition
$|F_b^2|=T_b$ we get eq (4.3).
\vfill
\eject
{\bf References}
\vskip 1pc
\item{1.} R.J. Glauber, in {\it Lectures in Theoretical Physics} 
 ed. W.E. Brittin (New York, 1959).
 M.M.Islam, in {\it Lectures in Theoretical Physics} vol 9B
 (New York, 1967).
\item{2.}   G. Calucci, D. Treleani Phys. Rev. {\bf D 41},3367   (1990)
        G.Calucci, D.Treleani, Phys. Rev. {\bf D 44},2746 (1991)
\item{3.}   G. Calucci, D. Treleani {\it in: Multiparticle dynamics 1998 - 
Delphi - Greece}   
\item{4.} H.M. Fried {\it Functional methods and models in QFT, ch.9 } M.I.T. 
 Press Cambridge, U.S.A. (1973)
\item{5.} H. Feshbach, in {\it Rendiconti S.I.F. - Course XXXVIII}
 (New York, 1967).
 R.L. Sugar and R. Blankenbecler Phys. Rev. {\bf 183}, 1387 (1969)
\item{6.} G. Calucci, R. Jengo, C. Rebbi, Nuovo Cimento {\bf 4A}, 330 (1971).
 G. Calucci, R. Jengo, Nuovo Cimento lett. {\bf 4}, 33 (1972).
\item{7.} J.D. Bjorken, in  ed. A.Giovannini,
S.Lupia, R.Ugoccioni (Singapore 1995)
\item{8.} O. Nachtmann, High-energy collisions and non perturbative QCD in {\it
  Lectures on QCD - Applications } ed. F.Lenz, H.Grie{\ss}hammer, D.Stoll, 
  (Springer 1997)
\item{9.} M. Drees and T. Han, Phys. Rev. Lett. {\bf 77}, 4142 (1996)
\item{10.} F. Abe et al., (CDF Collaboration), Phys. Rev. {\bf D 56}, 3811 
(1997)
\item{11.} M. Abramowitz and I. A. Stegun, {\it Handbook of Mathematical
   Functions}, Dover Publications, Inc., New York.
\item{12.} G. Calucci, D. Treleani,
                Phys. Rev. {\bf D 57}, 503 (1998)

\vfill   
\end
\bye